\RequirePackage{fix-cm}
\documentclass[twocolumn,epjc3]{svjour3}
\smartqed  
\RequirePackage{graphicx}
 \RequirePackage{mathptmx}      
\RequirePackage[numbers,sort&compress]{natbib}

\usepackage{amsmath}
\usepackage{amssymb}
\usepackage{url}

\newcommand\sch{Schwarzschild}
\newcommand\schbh{Schwarzschild black hole}

\begin{document}

\title{Strong deflection gravitational lensing by a modified Hayward black hole}


\author{Shan-Shan Zhao\thanksref{addr1,addr2}
        \and
        Yi Xie\thanksref{e2,addr1,addr2} 
}

\thankstext{e2}{e-mail: yixie@nju.edu.cn}

\institute{School of Astronomy and Space Science, Nanjing University, Nanjing 210023, China\label{addr1}
          \and
         Key Laboratory of Modern Astronomy and Astrophysics, Nanjing University, Ministry of Education, Nanjing 210093, China\label{addr2}
}

\date{Received: date / Accepted: date}

\maketitle

\begin{abstract}
	A modified Hayward black hole is a nonsingular black hole. It is proposed to form when the pressure generated by quantum gravity can stop matter's collapse as the matter reaches Planck density. Strong deflection gravitational lensing happening nearby its event horizon might provide some clues of these quantum effects in its central core. We investigate observables of the strong deflection lensing, including angular separations, brightness differences and time delays between its relativistic images, and estimate their values for the supermassive black hole in the Galactic center. We find that it is possible to distinguish the modified Hayward black hole from a \sch\ one, but it demands very high resolution beyond current stage.
\end{abstract}

\section{Introduction}


Einstein's general relativity (GR) breaks down at singularities. Quantum gravity plays an important role in the spacetime of a singularity and its surroundings, since energy density and curvature become very large in a tiny region \cite{Bojowald2005LRR11.8}. Therefore, sufficient knowledge of quantum gravity is necessary for studying the physics in the vicinity of a gravitational singularity. But the requested energy is too high to test quantum gravity effects by any Earth based experiments. Although we can lower the requested energy by testing the effects predicted by some specific models of quantum gravity, but there has not been any significant experimental results yet \cite{Liberati2011JPCS314.012007}. Astronomical observations on singularities might also test the quantum gravity and relate to two objects: the starting point of the Big Bang and the center of a black hole. 

It is believed that, in a quantum theory of gravity, the singularities in black holes under GR can be removed. A black hole without singularity is called a nonsingular black hole or a regular black hole which was first proposed by Bardeen \cite{Bardeen1968ConfGR5174}. Various models of static and spherically symmetric nonsingular black holes are reviewed in \cite{Ansoldi2008arXiv0802.0330,Frolov2016PRD94.104056} and rotating nonsingular black holes were also studied \cite{Bambi2013PLB721.329}. Lots of works \cite{Lunin2001NPB610.49,Balasubramanian2001PRD64.064011,Lunin2002NPB623.342,Maldacena2002JHEP12.055,Lunin2003NPB655.185,Mathur2005ForPh53.793,Rychkov2006JHEP01.063,Skenderis2007PRL98.071601,Kanitscheider2007JHEP04.023,Kanitscheider2007JHEP06.056,Bena2008LNP755.1,Mathur2009CQG26.224001,Susskind2012arXiv1208.3445,Almheiri2013JHEP02.062,Mathur2014JHEP01.34} in string theory and supergravity suggest that the singularity in the black hole might be replaced with a horizon sized ``fuzzball'', whose non-singular geometries are related to microstates of the black hole, so that the information loss paradox might be evaded.

A nonsingular black hole with a ``Plank star'' replacing the singularity was recently proposed \cite{Rovelli2014IJMPD23.1442026}. This scenario suggests that when matter collapses toward the center and reaches the Planck density, the pressure generated by quantum gravity becomes so large that could resist the collapse and finally form a bouncing instead of a singularity in the center of the black hole. In the process of Planck star forming, the explosion may produce detectable short gamma ray bursts \cite{Barrau2014PLB739.405,Barrau2016JCAP2.022} and fast radio bursts \cite{Christodoulou2016PRD94.084035}. The Hayward metric was originally chosen to describe the Planck star \cite{Rovelli2014IJMPD23.1442026}. It is an effective metric for a nonsingular black hole proposed by Hayward \cite{Hayward2006PRL96.031103} and has been well studied \cite{Abbas2014ApSS352.769,Halilsoy2014EPJC74.2796,Lin2013IJTP52.3771}.  In order to incorporate the 1-loop quantum corrections and a finite time delay between the center and infinity, a modified Hayward metric was then presented as the effective metric of the  Plank star \cite{Lorenzo2015GRG47.41}. Accretion and evaporation of the modified Hayward black hole was studied \cite{Debnath2015EPJC75.129} and its thermodynamics was also discussed \cite{Pourhassan2016EPJC76.145}. As another important and (possibly) observable aspect, investigation on gravitational lensing by the modified Hayward black hole is still absent in the literature, although the gravitational lensings by other kinds of nonsingular black holes have aroused a lot of concerns \cite{Eiroa2011CQG28.085008, Wei2015AHEP2015.454217, Schee2015JCAP06.048, AbdujabbarovPRD93.104004, Ghaffarnejad2016IJTP55.1492}. By studying its gravitational lensing effects, especially those in the strong gravitational field, we can have a better understanding of the modified Hayward black hole. 
Testing quantum gravity effects on a black hole would not be easy because an observer usually has to access the region extremely close to its event horizon. It was suggested that the quantum effects might occur at 7/6 \sch\ radius \cite{Haggard2016IJMPD25.1644021}. Gravitational lensing caused by strong field in the vicinity of a black hole can provide an opportunity for this purpose. Gravitational lensing in strong gravitational field is dramatically different from the one in a weak gravitational field. The unique phenomenon of the lensing effects in strong field was firstly discussed by Darwin in 1959 \cite{Darwin1959PRSLSA249.180}. A set of infinite discrete images, called relativistic images, will generate at the very close area on the two sides of the lens, due to photons winding several loops around the lens before reaching the detector, which belongs to  strong deflection gravitational lensing. The relativistic images are new observational phenomenon which can not be predicted by the classical gravitational lensing in a weak field. The strong deflection lensing by a \schbh\ have been well studied \cite{Luminet1979AA75.228,Ohanian1987AJP55.428,Nemiroff1993AJP61.619,Bozza2001GRG33.1535} and these effects caused by other static and symmetric black holes were also investigated \cite{Virbhadra1998AA337.1,Virbhadra2000PRD62.084003,Bozza2002PRD66.103001,Eiroa2002PRD66.024010,Bhadra2003PRD67.103009,Perlick2004PRD69.064017,Whisker2005PRD71.064004,Majumdar2005IJMPD14.1095,Eiroa2004PRD69.063004,Eiroa2005PRD71.083010,Keeton2006PRD73.044024,Eiroa2006PRD73.043002,Amore2006PRD73.083004,Amore2006PRD74.083004,Amore2007PRD75.083005,Iyer2007GRG39.1563,Mukherjee2007GRG39.583,Bozza2008PRD78.103005,Pal2008CQG25.045003,Virbhadra2009PRD79.083004,Chen2009PRD80.024036,Liu2010PRD81.124017,Bin-Nun2010PRD82.064009,Eiroa2012PRD86.083009,Eiroa2013PRD88.103007,Wei2015EPJC75.253,Wei2015EPJC75.331,Chen2015JCAP10.002,Man2015PRD92.024004,Huang2016EPJC76.594}. A more complicated scenario is the gravitational lensing in the strong field around a rotating black hole, which was widely discussed \cite{Bozza2003PRD67.103006,Vazquez2004NCimB119.489,Bozza2005PRD72.083003,Bozza2006PRD74.063001,Bozza2007PRD76.083008,Bozza2008PRD78.063014,Chen2010CQG27.225006,Chen2011PRD83.124019,Kraniotis2011CQG28.085021,Chen2012PRD85.124029,Kraniotis2014GRG46.1818,Ji2014JHEP03.089,Cunha2015PRL115.211102,Wang2016JCAP11.020}. These strong deflection lensing can be used to determine different black holes \cite{Bin-Nun2010PRD81.123011, Gyulchev2007PRD75.023006, Gyulchev2013PRD87.063005, Zhao2016JCAP07.007, Cavalcanti2016CQG33.215007}, naked singularities \cite{Gyulchev2008PRD78.083004,Sahu2012PRD86.063010,Sahu2013PRD88.103002,Virbhadra2002PRD65.103004} and wormholes \cite{Kuhfittig2014EPJC74.2818, Kuhfittig2015arXiv1501.06085,Nandi2006PRD74.024020,Tsukamoto2012PRD86.104062,Tsukamoto2016PRD94.124001} as well as test gravity \cite{Eiroa2014EPJC74.3171,Sotani2015PRD92.044052}. If the source of light has time signals, time delays between the relativistic images can also reveal some information about the lens \cite{Bozza2004GRG36.435,Virbhadra2008PRD77.124014,Man2014JCAP11.025,Lu2016EPJC76.357}. Reviews of strong deflection lensing can be found in \cite{Bozza2010GRG42.2269,Eiroa2012arXiv1212.4535}.

Direct observation on the gravitational lensing in strong field is still challenging since it requires a very high angular resolution. The most possible candidate to realize this observation is the supermassive black hole in the center of our Galaxy, called Sagittarius A* (Sgr A*). The apparent angular diameter (shadow) of Sgr A* is $\sim50$ microarcseond ($\mu$as) which is the largest one among all the known black holes \cite{Falcke2013CQG30.244003,Falcke2000ApJ528.L13}. First real image of Sgr A* will probably be detected by the Event Horizon Telescope \footnote{\url{http://www.eventhorizontelescope.org/}} which is an international sub-mm very long baseline interferometry (VLBI) network, and when it comes true, a new fundamental physics laboratory can be provided for testing the black hole physics as well as the gravity in a strong field regime \cite{Psaltis2008LRR11.9,Goddi2016arXiv1606.08879,Psaltis2015ApJ814.115,Broderick2014ApJ784.7}.

In this work, we will study the strong deflection gravitational lensing by the modified Hayward black hole. Its spacetime and domains of its model parameters are discussed in Sect. \ref{Sec2}. By using the strong deflection limit (SDL) method \cite{Bozza2002PRD66.103001}, we analytically describe the gravitational lensing in SDL in Sect. \ref{Sec3}. In Sect. \ref{Sec4}, the observables, including angular separations, brightness differences and time delays between the resulting relativistic images will be directly obtained. Then we estimate these observables for a modified Hayward black hole which has the same mass and distance of Sgr A* in Sect. \ref{Sec5}. Finally, conclusions and discussion are presented in Sect. \ref{Sec6}.

\section{Modified Hayward black hole}

\label{Sec2}

The modified Hayward metric discussed in this work was proposed in \cite{Lorenzo2015GRG47.41}. In order to fix its original two shortcomings, the original Hayward metric is modified by including the 1-loop quantum correction on the Newton potential \cite{Bjerrum-Bohr2003PRD67.084033,Bjerrum-Bohr2005PRD71.069903} and allowing for a non-trivial time dilation between a clock in the center of the black hole and a clock at infinity. Thus, the modified Hayward metric reads as (in the units of $G=c=1$) \cite{Lorenzo2015GRG47.41}
\begin{equation}
\label{eq:metric}
\mathrm{d}s^2=A(x)\mathrm{d}t^2-B(x)\mathrm{d}x^2-C(x)(\mathrm{d}\theta^2+\sin^2{\theta}\mathrm{d}\phi^2),
\end{equation}
where we take $2M$ as the measure of distances and set it to unity and 
\begin{eqnarray}
A(x)&=&\left(1-\frac{x^2}{l^2+x^3}\right) \left(1-\frac{\kappa  \lambda }{\lambda +2 \kappa  x^3}\right),\label{eq:metricA}\\
B(x)&=&\left(1-\frac{x^2}{l^2+x^3}\right)^{-1},\label{eq:metricB}\\
C(x)&=&x^2.\label{eq:metricC}
\end{eqnarray}
Here, $l$ is a parameter with a length dimension as the same one in the Hayward metric and it can introduce a repulsive force for the avoidance of the singularity \cite{Hayward2006PRL96.031103}; $\kappa$ and $\lambda$ are parameters used to modify the Hayward metric \cite{Lorenzo2015GRG47.41}. The parameter $\kappa\in[0, 1)$ relates to the time delay between $x=0$ and $x\to\infty$, and a larger time delay corresponds to a larger $\kappa$. When $\kappa=0$, there is no time delay between the center and the infinity, then the modified Hayward metric \eqref{eq:metric} degenerates to the Hayward metric. The parameter $\lambda$ indicates the strength of the 1-loop correction on the Newtonian gravitational potential \cite{Bjerrum-Bohr2003PRD67.084033,Bjerrum-Bohr2005PRD71.069903}.

The values of $l$ should be limited in a certain range to guarantee the existence of the event horizon(s), whose definition of $B^{-1}(x)=0$ gives a cubic equation of $x$ as 
\begin{equation}
\label{eq:1/B=0}	
x^3-x^2+l^2=0.
\end{equation}
According to Descartes' rule of signs, this equation has either two or none positive roots. In order to ensure that, the discriminant of this cubic equation $\Delta= -27 l^4 +4 l^2$ must be nonnegative so that we can have
\begin{equation}
l\le\frac{2}{3\sqrt{3}}\simeq0.385.
\end{equation}

In the following work on estimations of the strong deflection gravitational lensing observables, we set $l\in[0,2/(3\sqrt{3})]$ based on the above inequality and take $\kappa\in[0, 1)$ according to \cite{Lorenzo2015GRG47.41}. This domain of the parameters makes $A(x)>0$ and $B(x)>0$ for any $x$ outside the (outer) event horizon of the modified Hayward black hole.

It is worth emphasizing that, for the modified Hayward black hole, the geodesic rule is violated \textit{inside} the event horizon and such a violation is essential and critical to avoid all of the matter falling into the singularity. In fact, a quantum theory of gravity might induce interaction between electromagnetic and gravitational fields beyond the standard Einstein-Maxwell theory, such as one-loop vacuum polarization on the photon for quantum electrodynamics in which a photon might travel ``faster than light'' \cite{Drummond1980PRD22.343,Mankiewicz1989PRD40.2134,Khriplovich1995PLB346.251,Daniels1994NPB425.634,Shore1996NPB460.379,Daniels1996PLB367.75,Mohanty1998NPB526.501,Cho1997PRD56.6416,Cai1998NPB524.639,Shore2001NPB605.455,Shore2002NPB633.271}. It can make the worldline of a photon deviate from its geodesic. Strong deflection lensing of photons which do not follow geodesics were investigated \cite{Chen2015JCAP10.002,Lu2016EPJC76.357} and this deviation is characterized by an constant. In this work, we assume that the photons in the strong deflection gravitational lensing are following geodesics in the spacetime \textit{outside} the photon sphere of the modified Hayward black hole so that the SDL method can be applied. Although this assumption is still open to be tested, it would give a baseline for future works which take the violation of geodesic rule into account.

\section{Gravitational lensing under SDL}
\label{Sec3}

To analyze the strong deflection gravitational lensing by the modified Hayward black hole, we need two ingredients. One is a lens equation to define the geometrical relationships of the observer, the lens and the light source; the other is a deflection angle determined by the spacetime of the lens.

The lens equation given in \cite{Bozza2001GRG33.1535} is adopted in our work for its physical feasibility and widespread usage. There are three assumptions in applying this lens equation:
\begin{enumerate}
\item The spacetime is asymptotically flat at infinity;
\item The observer and the source are far from the lens;
\item The observer, lens and the source are nearly in alignment, and the source locates behind the lens.
\end{enumerate}
Under such assumptions, the lens equation could be written as \cite{Bozza2001GRG33.1535}
\begin{equation}
\label{eq:LE}
\beta=\theta-\frac{D_{\mathrm{LS}}}{D_{\mathrm{OS}}}\Delta\alpha_n,
\end{equation}
where $\Delta\alpha_n=\alpha(\theta)-2n\pi$ is the extra deflection angle of a photon winding $n$-loops and has a deflection angle $\alpha$; $\beta$ and $\theta$ are the angular separation between the source and the lens and the angular separation between the image and the lens; $D_{\mathrm{LS}}$ is the distance of the lens to the source and $D_{\mathrm{OS}}$ is the distance of the observer to the lens, both of them are the projection along the optical axis.

This asymptotically approximated lens equation might also be defined in other ways, which were summarized and discussed in \cite{Bozza2008PRD78.103005}. Some works have tried to define the lens equation for more general cases \cite{Frittelli1999PRD59.124001,Perlick2004LRR7.9,Perlick2004PRD69.064017,Frittelli2000PRD61.064021,Eiroa2004PRD69.063004,Eiroa2005PRD71.083010,Bozza2007PRD76.083008}. However, the lens equation (\ref{eq:LE}) is predominant, because its brief form make it possible to analyze the observational lensing effects of the modified Hayward black hole in a clear physical picture.

The deflection angle of a photon moving on the equatorial plane ($\theta=2\pi$) in a static and spherically symmetric spacetime is \cite{Virbhadra1998AA337.1,Weinberg1972Book}
\begin{equation}
\label{eq:dfagl}
\alpha(x_0)=-\pi+\int^{\infty}_{x_0}{2\sqrt{B(x)}\over{\sqrt{C(x)}\sqrt{{C(x)\over{C_0}}{A_0\over{A(x)}}-1}}}\mathrm{d}x,
\end{equation}
where $x_0$ represents the closest approach distance of the winding photon;  $A_0$ and $C_0$ are the values of $A(x)$ and $C(x)$ at $x=x_0$. The exact deflection angle of modified Hayward black holes could be found by substituting (\ref{eq:metricA})- (\ref{eq:metricC}) into (\ref{eq:dfagl}). 

The integral in (\ref{eq:dfagl}) has an approximated form in the weak deflection limit (WDL) by assuming the deflection angle is a small angle. However this classic WDL method fails in describing the deflection in the strong gravitational field. The divergence occurs when $x_0$ approaches the photon sphere \cite{Virbhadra2000PRD62.084003,Claudel2001JMP42.818}. An effective way to handle this problem is to expand the deflection angle near the photon sphere in the SDL \cite{Bozza2002PRD66.103001}. This method can provide an explicit physical picture and a straightforward connection to the observables, which will be discussed in Sect. \ref{Sec4}.

The radius of the photon sphere $x_m$ is defined as the largest positive root of the following equation \cite{Virbhadra2000PRD62.084003,Claudel2001JMP42.818}
\begin{equation}
\label{eq:phosph}
\frac{C'(x)}{C(x)}=\frac{A'(x)}{A(x)}.
\end{equation}
By assuming the closest distance $x_0$ is not too larger than $x_m$, the deflection angle can be expanded in the SDL as \cite{Bozza2002PRD66.103001}
\begin{equation}
\label{eq:alpha}
\alpha(\theta)=
-\bar{a}\log\left(\frac{\theta D_{OL}}{u_m}-1\right)+\bar{b}+\mathcal{O}(u-u_m).
\end{equation}
where $u$ is the impact parameter given by \cite{Virbhadra2000PRD62.084003,Weinberg1972Book}
\begin{equation}
u=\sqrt{\frac{C_0}{A_0}},
\end{equation}
and $u_m$ is the impact parameter evaluated at $x_m$. The impact parameter $u$ and the angular separation $\theta$ could be related by $u\approx \theta D_{\mathrm{OL}}$ when the lens equation (\ref{eq:LE}) is adopted. Meanwhile, $\bar{a}$ and $\bar{b}$ are the SDL coefficients and their expressions are \cite{Bozza2002PRD66.103001}
\begin{eqnarray}
\label{eq:bara}
\bar{a}&=&\frac{R_m}{2\sqrt{\beta_m}},\\
\bar{b}&=&-\pi+b_R+\bar{a}\ln\frac{2\beta_m}{A_m},
\end{eqnarray}
where we have
\begin{eqnarray}
\beta_m&=&{C_m(1-A_m)^2 \left(A_m C_m''-C_m A_m''\right)\over{2A_m^2C_m'^2}},\\
R_m&=&\frac{2(1-A_m)\sqrt{A_mB_m}}{A'_m\sqrt{C_m}},\\
b_R&=&\int^1_0 \left[\frac{2(1-A_m)\sqrt{A(z)B(z)}}{A'(z)C(z)\sqrt{\frac{A_m}{C_m}-\frac{A(z)}{C(z)}}}-\frac{R_m}{z\sqrt{\beta_m}}\right]\mathrm{d}z,
\end{eqnarray}
and $z$ is a new variable deduced from $x$ by
\begin{equation}
\label{eq:xtoz}
z={A(x)-A_m\over{1-A_m}}.
\end{equation}
All the quantities with subscript $m$ refer to their corresponding values at $x=x_m$; and $'$ and $''$ mean taking derivative against $x$ once and twice. Therefore, $\bar{a}$ and $\bar{b}$ can be directly determined by the metric (\ref{eq:metricA})- (\ref{eq:metricC}) of the modified Hayward black hole after fixing the model parameters $l$, $\kappa$ and $\lambda$.

Apart from the deflection angle, other important observables are the time delays between relativistic images if the light source is variable with time. The total travel time of a photon moving from the source to the observe is \cite{Bozza2004GRG36.435}
\begin{equation}
	T=\tilde{T}(x_0)-\int^\infty_{D_{\mathrm{OL}}}\left|\frac{\mathrm{d}t}{\mathrm{d}x}\right|\mathrm{d}x-\int^\infty_{D_{\mathrm{LS}}}\left|\frac{\mathrm{d}t}{\mathrm{d}x}\right|\mathrm{d}x,
\end{equation}
where $D_{\mathrm{LS}}$ is the projected distance between the source and the lens. The last two terms can be easily calculated since the photon is far from the lens. The first term $\tilde{T}(x_0)$ is \cite{Weinberg1972Book,Bozza2004GRG36.435,Sahu2013PRD88.103002}
\begin{equation}
\label{eq:dftime}
\tilde{T}(x_0)=\int^{\infty}_{x_0}{2\sqrt{B(x)C(x)A_0}\over{A(x)\sqrt{{C(x)\over{C_0}}{A_0\over{A(x)}}-1}}}\mathrm{d}x,
\end{equation}
which is divergent by $x_0\to x_m$, and it can also be manipulated with the method of the SDL as 
\begin{equation}
\label{eq:time}
\tilde{T}(u)=-\tilde{a}\ln{\left(\frac{u}{u_m}-1\right)}+\tilde{b}+\mathcal{O}(u-u_m),
\end{equation}
where $\tilde{a}$ and $\tilde{b}$ are coefficients in the SDL. For a spherically symmetric spacetime, it is found that $\tilde{a}= \bar{a}\, u_m$ \cite{Bozza2004GRG36.435}.

\section{Observables}

\label{Sec4}

Combining the lens equation (\ref{eq:LE}) with the deflection angle (\ref{eq:alpha}) and the time delay (\ref{eq:time}) in the SDL, we can find the observables of the strong deflection lensing, including angular separations, brightness differences and time delays between the relativistic images.

The angular separation between the lens and a $n$-loop relativistic image can be written as a combination of two parts \cite{Virbhadra2000PRD62.084003,Bozza2002PRD66.103001} 
\begin{equation}
\label{eq:theta0}
\theta_n=\theta_n^0+\Delta\theta_n,
\end{equation}
where $\theta^0_n$ is the angle corresponds to the relativistic image with the photon winds completely $2n\pi$ and $\Delta\theta_n$ is the extra part exceeding $2n\pi$. They have expressions as
\begin{eqnarray}
\label{eq:theta1}
\theta^0_n&=&\frac{u_m}{D_{\mathrm{OL}}}\left\{1+\exp\left[(\bar{b}-2n\pi)/\bar{a}\right]\right\},\\
\Delta\theta_n&=&\frac{u_m(\beta-\theta_n^0)D_{\mathrm{OS}}}{\bar{a}D_{\mathrm{LS}}D_{\mathrm{OL}}}\exp\left[(\bar{b}-2n\pi)/\bar{a}\right],
\end{eqnarray}
in which $\theta_n^0\gg\Delta\theta_n$. The brightness of the relativistic images will be magnified by the lensing. For the $n$-loop relativistic image, its magnification is \cite{Liebes1964PR133.835,Refsdal1964MNRAS128.295}
\begin{equation}
\mu_n=\frac{1}{(\beta/\theta)\partial\beta/\partial\theta}\bigg|_{\theta_n^0}.
\end{equation}

In practice, if the $1$-loop relativistic image can be distinguished from other inner packed ones, we can find three characteristic observables \cite{Bozza2002PRD66.103001}
\begin{eqnarray}
\label{eq:theta}
\theta _{\infty }&=&\frac{u_m}{D_{\mathrm{OL}}},\\
\label{eq:s}
s&=&\theta_1-\theta_\infty=\theta _{\infty } \exp \left(\frac{\bar{b}}{\bar{a}}-\frac{2 \pi }{\bar{a}}\right),\\
\label{eq:r}
r&=&2.5 \log _{10}\left(\frac{\mu_1}{\sum^\infty_{n=2}\mu_n}\right)=2.5 \log _{10}\left[\ \exp \left({2 \pi }/{\bar{a}}\right)\ \right].
\end{eqnarray}
Here, $\theta_\infty$ is the asymptotic position of the images with $n\rightarrow\infty$, i.e., angular radius of the photon sphere; $s$ is the angular separation between the $1$-loop relativistic image and the packed others ($n=2,\cdots,\infty$); and $r$ is the magnitude difference of their brightness. 

The time delay between different relativistic images can also be calculated. If we can distinguish the time signals of the $1$-loop relativistic image and the $2$-loop one, the delay between them $\Delta T_{2,1}$ is given by \cite{Bozza2004GRG36.435}
\begin{equation}
\label{eq:TD}
\Delta T_{2,1} = \Delta T_{2,1}^{0}+\Delta T_{2,1}^{1},
\end{equation}
where $\Delta T_{2,1}^{0}$ and $\Delta T_{2,1}^{1}$ are respectively the leading and correction term and they are
\begin{eqnarray}
\label{eq:TD0}
\Delta T_{2,1}^{0}&=&2 \pi  u_m,\\
\label{eq:TD1}
\Delta T_{2,1}^{1}&=&2\sqrt{\frac{B_m}{A_m}}\sqrt{\frac{u_m}{c_m}}\exp\left({\frac{\bar{b}}{\bar{a}}}\right)\nonumber\\
& & \quad \times \left[\exp\left({-\frac{\pi}{\bar{a}}}\right)-\exp\left({-\frac{2\pi}{\bar{a}}}\right)\right]
\end{eqnarray}
with 
\begin{equation}
\label{eq:cm}
c_m=\beta_m\sqrt{\frac{A_m}{C_m^3}}\frac{{C_m'}{}^2}{2(1-A_m)^2}.
\end{equation}

After taking the metric of the modified Hayward black hole and fixing the parameters $l$, $\kappa$ and $\lambda$, we can obtain the observables $\theta _{\infty }$, $s$, $r$ and $\Delta T_{2,1}$ for its strong deflection gravitational lensing.

\section{Estimations for Sgr A*}

\label{Sec5}

\begin{figure*}[htpb]
\begin{minipage}{1\textwidth}
\centering
\includegraphics[width=1\textwidth]{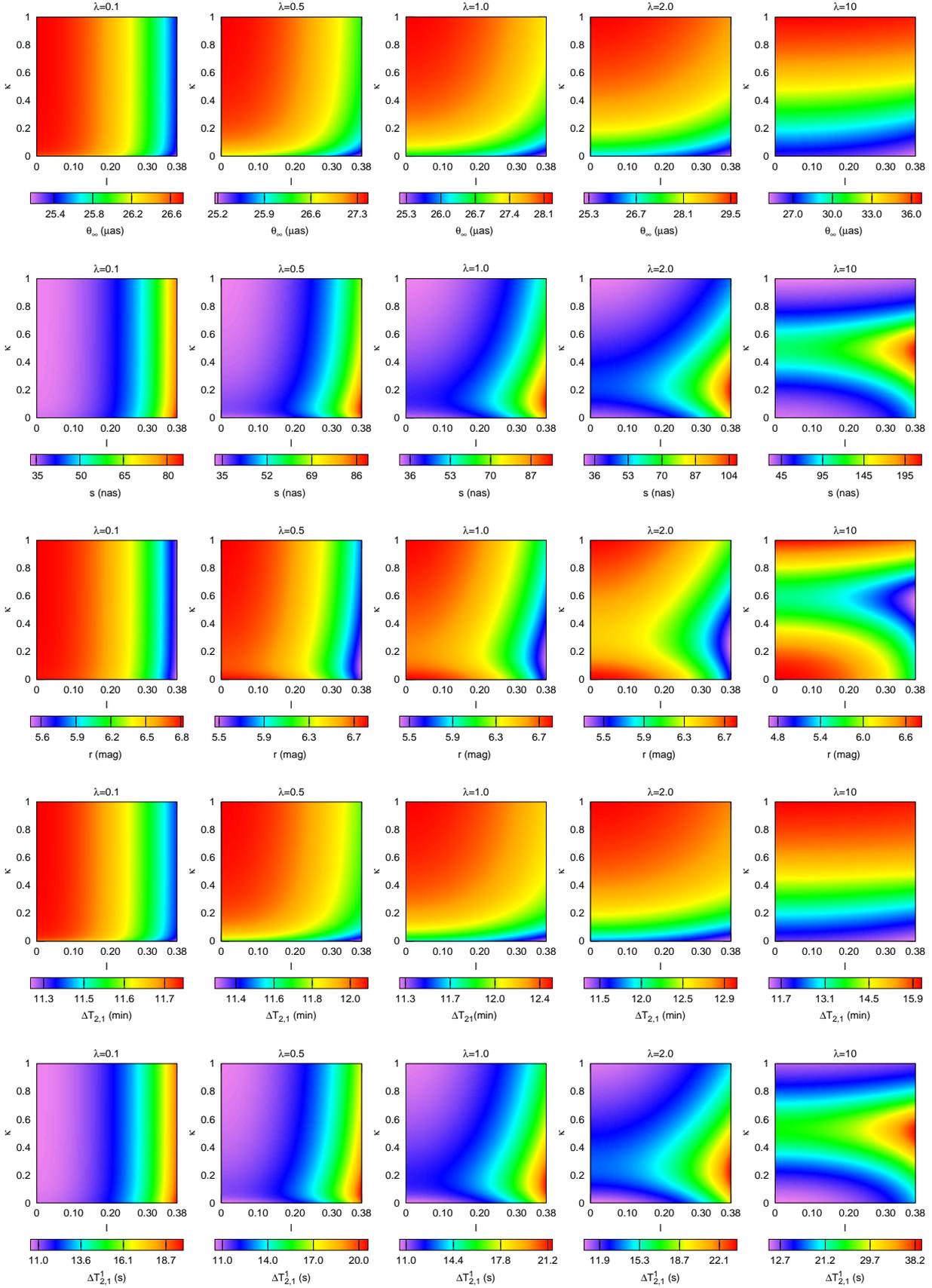}
\end{minipage}
\caption{Estimated observables of the strong deflection gravitational lensing caused by a modified Hayward black hole. From top to bottom panels, color-indexed $\theta_\infty$, $s$, $r$, $\Delta T_{2,1}$ and $\Delta T_{2,1}^1$ against $l$ and $\kappa$ are respectively presented by assuming such a black hole with the same distance and mass as Sgr A*: $D_{\mathrm{OL}}=8.33$ kpc and $M_\bullet=4.31\times10^6M_\odot$ \cite{Gillesse2009ApJ707.L114}. Specific values of $\lambda\in \{0.1, 0.5,1.0, 2.0, 10\}$ are chosen. For a comparison, a Schwazschild black hole ($\kappa=0$ and $l=0$) with the same distance and mass can have $\theta_\infty=26.54$ $\mu$as, $s=33.32$ nas, $r=6.822$ mag, $\Delta T_{2,1}=11.72$ min and $\Delta T_{2,1}^1=10.56$ s.}
\label{fig1}
\end{figure*}

In this section, all of the observables obtained by the SDL method will be estimated by taking the supermassive black hole in the Galactic center, Sgr A*, as an example of the modified Hayward black hole. With the coefficients of the modified Hayward metric (\ref{eq:metricA}) to (\ref{eq:metricC}) and specifying values for the parameters $l$, $\kappa$ and $\lambda$, we can estimate the observables according to equations (\ref{eq:theta}) to (\ref{eq:TD1}) in which the SDL coefficients are calculated numerically. As shown in equations (\ref{eq:theta}) and (\ref{eq:TD0}), both the radius of the photon sphere $\theta_\infty$ and the leading term of the time delay $\Delta T_{2,1}^0$ are directly proportional to $u_m$ so that new information provided by the time delay would be contributed from the correction term $\Delta T_{2,1}^1$ although $\Delta T_{2,1}^1 \ll \Delta T_{2,1}^0$. All of these observables are represented in color-indexed Fig. \ref{fig1}.

Figure \ref{fig1} shows that, when $\lambda$ and $\kappa$ are fixed, the increase of $l$ can make $\theta_\infty$, $r$ and $\Delta T_{2,1}$ decrease but cause $s$ and $\Delta T_{2,1}^1$ to grow, which physically means shrinking of the photon sphere, stretching the gap between the $1$-loop relativistic image and the packed others, weakening their brightness difference, lessening the time delay between them and enlarging the correction term in the time delay. When a bigger value of $\lambda$ is taken, all of the observables will become less sensitive to the variation of $l$ for a given $\kappa$. If we fix $l$ and $\lambda$, Fig. \ref{fig1} demonstrates that the role of $\kappa$ will be played in a more complex way. In the case of $\lambda=0.1$, the augmentation of $\kappa$ can barely affect the values of all the observables. However, for a bigger $\lambda$, while the growth of $\kappa$ causes $\theta_\infty$ and $\Delta T_{2,1}$ to increase monotonically, it can make $s$ and $\Delta T_{2,1}^1$ increase first and then decrease and their locally maxima can be found at a larger $l$. A similar behavior happens for $r$ but the consequence of a growing $\kappa$ make it decrease and then increase, which is opposite to the tendency of $s$ and $\Delta T_{2,1}^1$. It is also clear that the patterns of these color-indexed figures are strongly affected by the values of $\lambda$, especially in the cases of $s$, $r$ and $\Delta T_{2,1}^1$. In principle, $\lambda$ can enhance the maximum values of the observables given by the domain of $l$ and $\kappa$, except for $r$.

More specifically, $\theta_\infty$ can vary widely, ranging from 25.2 to $36.8$ $\mu$as in Fig. \ref{fig1}. It has a theoretically lower limit of 25.11 $\mu$as when $\kappa=0$, $l\to2/(3\sqrt{3})$ and $\lambda\to0$. The value $\theta_\infty=26.54$ $\mu$as, which corresponds to angular radius of the photon sphere of a \schbh\ with the same mass and distance, is also permitted. It means that the measurement on $\theta_\infty$ itself cannot distinguish the modified Hayward black hole from the \sch\ one. Other observables are needed for this purpose. We find that, based on Fig. \ref{fig1}, the angular separation $s$ ranges from about 30 to about 200 nanoarcsecond (nas). For the smallest $s$, if the angular resolution can reach about 10 nas or better, which is far beyond current capabilities, the $1$-loop relativistic image and the packed others will be able to separated so that it is possible to measure their brightness difference and the time delay. According to Fig. \ref{fig1}, it is found that $r$ ranges from 4.7 to 6.8 mag; $\Delta T_{2,1}$ can reach from 11.3 to 16.2 minutes and its correction term can have values of tens seconds, which means $s$ and $\Delta T_{2,1}$ and its correction $\Delta T_{2,1}^1$ might be accessible under such an extremely high resolution. These additional constraint imposed by $s$, $r$ and $\Delta T_{2,1}$ can be helpful for pinning down the modified Hayward black. 

The modified Hayward black hole we have discussed above is a non-rotating one. An astrophysical black hole is very likely spinning. In order to describe the spacetime of a rotating modified Hayward black hole and the gravitational lensing happening in its vicinity in a self-consistent way, its metric is indispensably needed. Although some rotating regular black holes are known \cite{Bambi2013PLB721.329}, it was found \cite{Lorenzo2016GRG48.31} that the metric of a rotating modified Hayward black hole is not unique and has no closed causal curves for any positive radial coordinates. As suggested in \cite{Lorenzo2016GRG48.31}, studying its geodesics equation might be able to provide helpful insights on the properties of this non-singular rotating metric. Investigations on gravitational lensing by the rotating modified Hayward black hole should be proceeded with caution. Nevertheless, based on the work on strong deflection lensing by a Kerr black hole \cite{Bozza2003PRD67.103006}, we can intuitively expect that the angular momentum of a modified Hayward black hole would also drift its caustics away from the optical axis, make the caustic with a finite extension and cause only one image visible instead of two sets of relativistic images. Direct imaging might not be able to independently and simultaneously determine the spin and its inclination relative to the observer, whose degeneracy would be broken by observing its high order effects \cite{Bozza2005PRD72.083003,Bozza2006PRD74.063001}. A detailed study on the strong deflection lensing by a rotating modified Hayward black hole will be left for our future work given the fact that knowledge of such a rotating metric is still limited for now. 

Accretion flow and its emission around Sgr A* will significantly affect the observations in the wavelength of millimeter on the angular radius of the photon sphere (``shadow'') and the relativistic images. However, since current understanding of accretion physics is still incomplete and the emission from Sgr A* is expected to have time-dependent properties, it is not feasible to model and predict the details of the brightness profile of the image of the accretion flow \cite{Jaroszynski1997A&A326.419,Falcke2000ApJ528.L13,Broderick2009ApJ697.45,Broderick2011ApJ735.110,Dexter2009ApJ703.L142,Dexter2010ApJ717.1092,Moscibrodzka2009ApJ706.497,Moscibrodzka2013A&A559.L3,Moscibrodzka2014A&A570.A7,Chan2015ApJ799.1,Chan2015ApJ812.103}. In principle, the boundary of the shadow is surrounded by a bright ring, whose width is about a few to tens $\mu$as (see Figure 4 in \cite{Psaltis2015ApJ814.115} based on simulation in \cite{Chan2015ApJ799.1}). Because the angular separation of the outermost relativistic images is about 0.2 $\mu$as even in the optimistic cases for a modified Hayward black hole, the relativistic images will merge and mix with the emission of the flow. The possibility and methodology for detecting relativistic images in such a circumstance are still open problems.

In practice, the parameters $l$, $\kappa$ and $\lambda$ of the modified Hayward black hole will not be uniquely determined by measuring a single observable (see Fig. \ref{fig1}). In order to break their degeneracy, three different type observables, such as $\theta_{\infty}$, $s$ and $r$, are required at least. But, even these observations are available, determination of the values and their uncertainties of the parameters will not be trivial considering that there are a lot of uncertainly astrophysical factors contributing to these observation. Some model-independent and sophisticated inference methods \cite{Psaltis2015ApJ814.115,Johannsen2016PRL116.031101,Kim2016ApJ832.156} need to be employed.

\section{Conclusions and Discussion}

\label{Sec6}

A modified Hayward black hole is a nonsingular black hole with a Planck star in its center supported by the pressure due to quantum gravity. We investigate its strong deflection gravitational lensing by the method of SDL, in which the SDL coefficients are calculated numerically, and obtain its observables, including the radius of the photon sphere as well as the angular separations, the brightness differences and the time delays between relativistic images. In order to discuss its preliminary detectability, these observables are estimated by taking Sgr A* as a modified Hayward black hole. We find that if the photon sphere can only be measured and none of relativistic images can be resolved, the modified Hayward black hole can possess an identical $\theta_\infty$ of a \schbh, although they have totally different spacetime in their center. If the angular resolution can reach 10 nas or better, the $1$-loop relativistic image can be separated from the packed ones so that other observables, including the angular separation $s$ and the brightness differences $r$ and (possible) time delay $\Delta T_{2,1}$ between these relativistic images, will be helpful for constraining the modified Hayward black hole.

However, such an extremely high resolution is far beyond capabilities of current technologies. In fact, other fundamental factors, such as the accretion flow and the scattering by interstellar electrons \cite{Psaltis2015ApJ814.115,Broderick2014ApJ784.7,Fish2014ApJ795.134}, will also complicate tests of black hole physics by the upcoming VLBI image of Sgr A*. Nevertheless, astronomical observations on gravitational lensing in strong field regime still provides a possible opportunity in the future for searching and detecting a modified Hayward black hole.


\begin{acknowledgements}
This work is funded by the National Natural Science Foundation of China (Grant No. 11573015).
\end{acknowledgements}



\bibliographystyle{spphys}

\bibliography{Refs20170418}

\end{document}